\documentstyle[prl,twocolumn,aps]{revtex}

\newcommand{\Nn}{{\cal N}}
\newcommand{\NN}{{\bf N}}
\newcommand{\JV}{\vec{J}}

\newcommand{\XV}{\vec{X}}

\begin{document}
\tolerance 50000
\draft
\twocolumn[\hsize\textwidth\columnwidth\hsize\csname @twocolumnfalse\endcsname

\title{Spectrum and diffusion for a class of tight-binding models on
hypercubes}

\author{Julien Vidal$^{1}$, R\'{e}my Mosseri$^{1}$ and Jean Bellissard$^{2,3}$}
\address{$^{1}${\it Groupe de Physique des Solides,}
{\it Universit\'{e}s P. et M. Curie P6 et D. Diderot P7,}\\
{\it Tour 23, 2 place Jussieu, 75251 Paris}
Cedex 05\\}

\address{$^{2}$ \it Institut Universitaire de France}

\address{$^{3}${\it Laboratoire de Physique Quantique, UMR5626 associ\'ee au
CNRS, IRSAMC, Universit\'e Paul Sabatier,\\
 118, route de Narbonne, F-31062 Toulouse Cedex 4, France}}

\maketitle

\begin{abstract}
We propose a class of exactly solvable anisotropic tight-binding models on an
infinite-dimensional hypercube.
The energy spectrum is analytically computed and is shown to be fractal and/or
absolutely continuous according to the value hopping parameters. In both cases,
the spectral and diffusion exponents are derived.
The main result is that, even if the spectrum is absolutely continuous, the
diffusion exponent for the wave packet may be anything between 0 and 1
depending upon the class
of models.
\end{abstract}

\vskip2pc]

\vspace{.5cm}

The interplay between the energy spectrum and the diffusion process for quantum
systems is still an open question. Indeed, recent studies have shown that the
spreading of a wavepacket is determined by several exponents that only depend
on the nature of the spectrum (absolutely continuous, singular continuous, pure
point, or any mixture)\cite{KPG,KKKG,Guarneri}. In particular, for
quasiperiodic
systems, such as quasicrystals (QC), it has been shown that an anomalous
behaviour of the wavepacket spreading occurs when the  spectrum is singular
continuous\cite{Zhong} (see below).

\noindent In such systems, scaling laws occur at many levels. We will
concentrate upon two classes of scaling exponents due to their importance for
electronic and transport properties. The first one concerns {\em spectral
exponents}, defined as $\int_{E-\delta E}^{E+\delta E} d\Nn(E') \sim \delta E
^{\alpha (E)}$ as $\delta E \rightarrow 0$, where $\Nn$ is the local density of
states (LDOS). An absolutely continuous spectrum in some interval $I$ of energy
implies $\alpha(E) =1$ for $E\in I$ and $E$ in the spectrum. A pure point
spectrum
in $I$, namely the
LDOS is a sum of Dirac peaks in $I$, implies $\alpha (E)=0$ on the part of the
spectrum contained in $I$. Finally, if $0 < \alpha (E) <1 $ for $E$ in some
part of the spectrum implies singular continuous spectrum there. This is
actually what happens for $1D$ QC \cite{Suto,BTS} such as the
Fibonacci chain. Such singular continuous spectrum occurs also for $1D$ chains
with a potential given by a substitution sequence
\cite{Be88,BBG,BovGhe,SimKnill}. The problem of $2D$ electrons on a square
lattice
under magnetic field that can be mapped onto a $1D$ Harper equation
\cite{Hofstadter}
also display a singular continuous spectrum \cite{XX} for incommensurate
fluxes, and the
spectral exponents have been computed numerically \cite{XX2}. A number of non
rigorous results suggesting singular spectra were obtained for models on higher
dimensional QC such as the {\em labyrinth model} \cite{Sire2}, the
{\em octagonal tiling} \cite{Benza}, in a regime of small hopping. At last
let us
also mention recent studies on Jacobi matrices associated to {\em iterated
functions systems} (IFS) \cite{GuaSchuba,SchuBar} generalizing Julia sets
\cite{BBM,Barnsley} leading to the calculation of the spectral exponents.

\noindent The next class concerns {\em diffusion exponents}, given by
$L_{E}(t) \sim t^{\beta (E)}$ as $t\rightarrow \infty$ where $L_{E}(t)$
characterizes the spreading of a typical wave packet of energy $E$ after time
$t$ \cite{BeSchuba} (see eq.~(\ref{definition}) for a definition of the
exponents). {\em Ballistic} motion such as the electronic motion in a perfect
crystal, corresponds to $\beta =1$. {\em Strong localization} is defined by
demanding that $\sup_{t >0} L_E(t) < \infty$ implying $\beta (E)=0$. The {\em
weak localization} regime appearing in weakly disordered metals generally
corresponds to $\beta =1/2$, namely to the case for which the quantum evolution
mimicks a classical diffusion such as a Brownian motion. Such a behaviour
of the
quantum diffusion also occurs in the Random Phase Approximation (RPA)
\cite{BeSchuba}. If $\beta$ takes on values different from $0,1/2,1$ we will
speak of {\em anomalous diffusion}. In QC's for instance, numerical results
show
that $\beta$ may be anything between $0$ and $1$ and is model dependent
\cite{Zhong,Sire3}. For realistic QC's, such as $i\mbox{\rm -}AlCuCo$, an
{\it ab
initio} calculation leads to $\beta = .375$ at the Fermi level \cite{FujiRo}.

Spectral and diffusion exponents are related through the {\em Guarneri
inequality} \cite{Guarneri},
%%%%%%%%%%%%%%%
\begin{equation}
\label{guarneri}
\beta (E) \geq \frac{\alpha (E)}{d}
\mbox{ , }
\end{equation}
%%%%%%%%%%%%%%%
\noindent where $d$ is the dimension of the system. This inequality implies
that
for absolutely continuous spectrum, ballistic motion always occurs in $1D$
(see however \cite{remark,Simon2}), whereas in higher dimension $\beta \geq
1/d$. One of the main
question addressed recently in this respect is whether $\beta$ can be computed
more directly from $\alpha$. It seems that this is indeed the case for Jacobi
matrices \cite{GuaSchuba,SchuBar} namely $1D$ chains with nearest neighbour
interactions. One important result of this paper is precisely to show the
opposite in the extreme case for which $d=\infty$, namely that there is no
relation whatsoever between the $\alpha$'s and the $\beta$'s. This gives a
negative answer to the question raised by Lebowitz (see \cite{Simon2}).
This is because the spectral exponents characterize only the spectral measure
(the LDOS) of the Hamiltonian $H$, independently of any other type of
observables. On the other hand, the diffusion exponents involves the
interplay between the
Hamiltonian (through the quantum evolution) and the {\em position operator}
$\XV$, or even better, the {\em current} $\JV = i [H,\XV]/\hbar$. The
link between the diffusion exponent $\beta$ and the pair
$(H,\JV)$ is still not  precisely established, even though it has been
related to the {\em
current-current correlation function} \cite{Pastur,BeSchuba}. However, Jacobi
matrices are very special since the position operator is defined by mean of the
orthogonal polynomials associated to the spectral measure (the LDOS), so that
 $\beta$ is defined through purely spectral properties.

In this letter, we consider a family of anisotropic tight-binding models in an
infinite dimensional hypercubic structure and show that depending upon the
explicit form of the hopping parameters, it is possible to shift from an
absolutely continuous spectrum to a singular con\-tinuous spectrum.
Moreover, we are able to adjust the fractal dimension of this spectrum,
fine-tuning a single parameter that drives the transition.
In addition, we show that depending on the  hopping term law, one can face an
absolutely continuous spectrum and a somewhat anomalous diffusion for which the
mean square deplacement $L(t)$ can either scale as $\log t$ or as $t^{\alpha}$
with $0<\alpha<1$. We first introduce some mathematical tools that are useful
for a careful analysis of the structure we are dealing with. We then
characterize
the energy spectrum for different types of tight-binding hamil\-to\-ni\-ans and
discuss the nature of their spectral measures. Fi\-nal\-ly, we compute the
autocorrelation function and the mean square displacement of a wavepacket for
the different models.\\

A $d$-dimensional hypercube $\Delta_d$ is the set of vertices of a cube of size
1 in an $d$-dimensional space. The infinite dimensional hypercube $\Delta$ is
defined by: $\Delta=\bigcup_{d>1} \Delta_d$. It can therefore be seen as
the set
of sequences $\varepsilon = (\varepsilon_k)_{k=0}^{\infty}$ where
$\varepsilon_k
\in \{ 0,1\}$ and $\varepsilon_k =0$ for all but a finite number of $k$'s. We
endow the set $\{ 0,1\}$ with the group structure given by the addition {\it
modulo} 2, so that $\Delta$ becomes a discrete countable group for the
coordinatewise addition. It is also convenient to introduce its dual group
$\cal
B$ ($\cal B$ stands for Brillouin), which is the counterpart of the
quasimomentum space in a perfect crystal. $\cal B$ can be described as the set
of all sequences $\sigma = (\sigma_k)_{k=0}^{\infty}$ with $\sigma_k = \pm 1$.
$\cal B$ is a compact abelian group with the pointwise multiplication and
the product
topology. The duality between $\Delta$ and $\cal B$ is given by the characters:
%%%%%%%%%%%%%%%
\begin{equation}
\label{character}
\forall \sigma\in{\cal B}, \forall\varepsilon\in{\Delta}, \hspace{2.ex}
\chi_{\sigma} (\varepsilon) =
    \prod_{k=0}^{\infty}
     \sigma_k^{\varepsilon_k}
\mbox{ . }
\end{equation}
%%%%%%%%%%%%%%%

\noindent In this formula, the product is finite by construction. Moreover, $
\chi_{\sigma} (\varepsilon + \varepsilon ')=\chi_{\sigma} (\varepsilon)
\chi_{\sigma} (\varepsilon ')$ and  $ \chi_{\sigma +\sigma '}
(\varepsilon)=\chi_{\sigma} (\varepsilon) \chi_{\sigma '} (\varepsilon)$ .
These characters play the role of the Bloch phase $\exp{(i ka)}$ in a
crystal, where
$a$ is a period of the translation group, and $k$ is a quasimomentum.
While on $\Delta$ the Haar measure is the counting one, the integral of a
continuous function $f$ on $\cal B$ is defined as:
%%%%%%%%%%%%%%%
\begin{equation}
\label{haarB}
\int_{\cal B}  \
  d\sigma f(\sigma) =
  \lim_{K\rightarrow \infty}
   \frac{1}{2^K}
    \sum_{\sigma_0 = \pm1, \cdots , \sigma_K = \pm1}
     f(\sigma)
\mbox{ . }
\end{equation}
%%%%%%%%%%%%%%%

The Hilbert space of physical states is ${\cal H}= \ell ^2(\Delta)$, namely the
set of sequences $\psi (\varepsilon)$ indexed by $\Delta$ (the wave functions),
such that:

%%%%%%%%%%%%%%%
\begin{equation}
|| \psi ||^2 =
 \sum_{\varepsilon \in \Delta}
  |\psi (\varepsilon)|^2
   < +\infty
\mbox{ . }
\end{equation}
%%%%%%%%%%%%%%%

\noindent A canonical orthonormal basis is provided by the states
$|\varepsilon>$
vanishing everywhere but on the "site" $\varepsilon$. The Fourier transform
of a
wave function $\psi \in {\cal H}$ is the function on $\cal B$ formally defined
by:

%%%%%%%%%%%%%%%
\begin{equation}
{\cal F}\psi (\sigma) =
 \sum_{\varepsilon \in \Delta}
  \chi_{\sigma}(\varepsilon)
   \psi (\varepsilon)
\mbox{ . }
\end{equation}
%%%%%%%%%%%%%%%

\noindent This function actually belongs to $L^2 (\cal B)$, namely it is square
integrable on $\cal B$ (with respect to the Haar measure) and the Parseval
identity holds true namely $||\psi ||^2 = ||{\cal F}\psi||^2 =\int_{\cal B}
d\sigma |{\cal F}\psi (\sigma)|^2$. Therefore we get two unitarily equivalent
representations of the Hilbert space of states.

The translation operators $T(a), (a\in\Delta)$ are acting on ${\cal H}$ as
follows:

%%%%%%%%%%%%%%%
\begin{equation}
T(a)\psi (\varepsilon) =
   \psi (\varepsilon -a)
\mbox{ . }
\end{equation}
%%%%%%%%%%%%%%%

\noindent Equivalently, $T(a) |\varepsilon > = |\varepsilon -a >$. Note that $a
= -a$ in $\Delta$ due to the addition {\it modulo} 2, so that $T(a)^2 = 1,
\forall
a \in \Delta$. In addition $T(a)=T(a)^{\dag}$ as can be easily checked, so that
there is a infinite set of mutually commuting unitary and self-adjoint
operators. The spectrum of such operators is made of two eigenvalues (with
infinite multiplicities) namely $\pm 1$. Through a Fourier transform, $T(a)$
becomes the operator of multiplication by $\chi_{\sigma}(a)$.
Par\-ti\-cu\-lar\-ly, if $a=e_k$, where $e_k$ is the sequence in $\Delta$ with
all coordinates vanishing except the $k$-th one, $T_k =T(e_k)$ becomes simply
the operator of multiplication by $\sigma_k$.\\

We consider the following class of tight-binding hamiltonians on ${\cal H}$:

%%%%%%%%%%%%%%%
\begin{equation}
H =
 \sum_{k=0}^{\infty}
  t_k T_k
\mbox{ . }
\end{equation}
%%%%%%%%%%%%%%%

\noindent In order that $H$ be self-adjoint we need $t_k \in {\bf R}$. By a
simple unitary tranformation, one can choose $t_k \geq 0$. The coefficient
$t_k$
denote the "transfer" or "hopping" term in the $k$-th direction. $H$ is
bounded
if and only if $\sum t_k < +\infty$. It is self-adjoint (but not necessarily
bounded) if $\sum t_k^2 < +\infty$. In what follows, we will assume that this
latter condition holds. By Fourier tranform, $H$ becomes the operator of
multiplication by $E(\sigma)$ where $E$ is called the {\em band function}
and is
given by:

%%%%%%%%%%%%%%%
\begin{equation}
E(\sigma) =
 \sum_{k=0}^{\infty}
 \sigma_k  t_k
\mbox{ . }
\end{equation}
%%%%%%%%%%%%%%%

\noindent This function is real and square integrable on $\cal B$ with
${\cal L}^2$ norm:

%%%%%%%%%%%%%%%
\begin{equation}
\int_{\cal B} d\sigma E(\sigma)^2 = \sum_{k=0}^{\infty}t_k^2
\mbox{ . }
\end{equation}
%%%%%%%%%%%%%%%

\noindent The spectrum of $H$ (resp. its spectral measure), is then given by
the image of $\cal B$ in ${\bf R}$ (resp. of the measure $d\sigma $) under the
function $E$. Note that if $H$ is bounded $E$ is continuous with
$||H|| =\sup_{\sigma \in {\cal B}} |E(\sigma)| =  \sum_{k=0}^{\infty} t_k$.

\vspace{.1cm}

Spectral properties can be studied through the autocorrelation function:

%%%%%%%%%%%%%%%%%%%%%%%%%%%%%%
\begin{equation}
P(s) = |<0|e^{i sH}|0>|^2
  = \left(
      \int_{\bf R} d\mu (E) e^{i sE}
    \right)^2
\mbox{ , }
\end{equation}
%%%%%%%%%%%%%%%

\noindent where $|0>$ denotes an origin site where we initially localize a
wavepacket and $\mu$ the corresponding spectral measure. Note that the
translation invariance of $H$ allows us to choose any site of $\Delta$ as
initial condition. If  $P$ is integrable over $\bf R$, then $\mu$ is absolutely
continuous (the converse may not be true). One can alternatively use the
temporal correlation function:

%%%%%%%%%%%%%%%
\begin{equation}
C(t) =
 {1\over t} \int_{0}^t ds P(s)
\mbox{ , }
\end{equation}
%%%%%%%%%%%%%%%%%%%%%%%%%%%%%%

\noindent that is the time-averaged version of $P$. The spectral measure is
purely continuous (singular or absolutely continuous) if and only if $C(t)
\rightarrow 0$ as $t\rightarrow \infty$ (Wiener criterion).

\noindent An elementary computation using (\ref{haarB}) leads to:

%%%%%%%%%%%%%%%
\begin{equation}
\label{piofess}
P(s)=\prod_{k=0}^{\infty}\cos^2(st_k)
\mbox{ . }
\end{equation}
%%%%%%%%%%%%%%%

\noindent This infinite product converges since $\sum t_k^2 <+\infty$.

\noindent We define the position operator as follows. For $k\in {\bf N}$,
$X_k$ denotes
the operator of multiplication by $\varepsilon_k$ in ${\cal H}$. It commutes
with $T_l$ for $l\neq k$ whereas $T_k X_k T_k^{-1} = {\bf 1} -X_k$, and since
$T_k^2={\bf 1}$ it follows that:
\begin{equation}
e^{i s T_k} =
   \cos{s} +i T_k \sin{s}
\mbox{ . }
\end{equation}

\noindent Let us define the mean square displacement by:

%%%%%%%%%%%%%%%
\begin{equation}
L_E^2(s) =
 \sum_{k=0}^{\infty}
  <\varphi |
   \left(
      X_k (s)-X_k(0)
   \right)^2
    |\varphi >
\mbox{ , }
\end{equation}
%%%%%%%%%%%%%%%

\noindent where $X_k(s) = \exp{(i s H)}X_k \exp{(-i s H)}$
denotes the Heisenberg representation of $X_k$ and $|\varphi>$ is an initial
state with energy close to $E$. This expression does not depend upon the
explicit choice of
$\varphi$ as it turns out. Using the previous relations, one gets for any $E$:

%%%%%%%%%%%%%%%
\begin{equation}
\label{loclength}
L_E^2(s) = L^2(s) =
 \sum_{k=0}^{\infty}
  \sin^2(s t_k)
\mbox{ . }
\end{equation}
%%%%%%%%%%%%%%%

\noindent The diffusion exponent $\beta$ is given by $L(s) \sim
s^{\beta}$ as $s\rightarrow \infty$. It does not depend on $E$. A rigourous
way to define a power
law asymptotic behaviour is given as follows (see  \cite{BeSchuba,BZM} for
more details):
a function $f$ of a real variable $s$ behaves as $s^{\beta}$ when
$s\rightarrow \infty$
if:
%%%%%%%%%%%%%%%
\begin{equation}
\label{definition}
\int_c^{\infty} {ds \over s^{1+b}} f(s)
\mbox{ , }
 \hspace{3.ex} c>0
\mbox{ , }
\end{equation}
%%%%%%%%%%%%%%%
converges for $b>\beta$ and diverges for $b < \beta $. In addition, if the
function
$f$ can be written as a series:

%%%%%%%%%%%%%%%
\begin{equation}
\label{serie}
f(s)=\sum_{k=0}^{\infty} F(st_k)
\mbox{ , }
\end{equation}
%%%%%%%%%%%%%%%
where $F$ is a positive bounded real function, such that $F(x)=O(x^2)$ for
$x\sim 0$, and
$(t_k)_{k\in \NN}$  a set of positive number such that $\sum_{k=0}^{\infty}
t_k^2 <\infty$,
then the exponent $\beta$ is given by:
%%%%%%%%%%%%%%%
\begin{equation}
\label{beta}
\beta=\inf{ \{ b\in {\bf R_+}; \sum_{k=0}^{\infty} t_k^b <\infty \} }
\mbox{ . }
\end{equation}
%%%%%%%%%%%%%%%
It is clear that this definition is particularly convenient for our purpose
since $L^2$
is exactly of the form (\ref{serie}).

\noindent The closed forms obtained for the three observables previously
discusssed (energy, autocorrelation function, mean square displacement), allows
us to study the spectrum and the quantum diffusion for various classes of
hopping terms.

%%%%%%%%%%%%%%%%%%%%%%%%%%%%%%%%%%%%%%%%%%%%%%%%%%%%%%%%%%%%%%%%%%%%%%%%%%%%
%%%%%%%%%%%%%%%  Etude des differents modeles %%%%%%%%%%%%%%%%%%%%%%%%%%%%%%
%%%%%%%%%%%%%%%%%%%%%%%%%%%%%%%%%%%%%%%%%%%%%%%%%%%%%%%%%%%%%%%%%%%%%%%%%%%%

A first interesting class of models consists in choosing an algebraic
scaling of the
hopping parameters $t_k \sim k^{-\gamma}$, namely $\lim_{k\rightarrow
\infty}k^{\gamma}
 t_k =t $, with $\gamma >1/2$. In this case, the spectrum is bounded if
$\gamma > 1$
whereas it is unbounded if $\gamma \leq 1$. Moreover, one can prove (see
Appendix) that

%%%%%%%%%%%%%%%
\begin{equation}
\label{Pest}
P(s) \leq c_1 e^{-c_2 s^{1/\gamma}}
\mbox{ , }
\end{equation}
%%%%%%%%%%%%%%%

\noindent where $c_1,c_2$ are two positive constants. This shows that the
spectral measure is always absolutely continuous and also infinitely
differentiable. This also implies that the correlation function decays as
$1/t$.
In addition, according to expression (\ref{beta}), it is obvious that:

%%%%%%%%%%%%%%%
\begin{equation}
L^2(s) \sim \, s^{1/\gamma}
\mbox{ . }
\end{equation}
%%%%%%%%%%%%%%%

\noindent Hence, the diffusion exponent is $\beta = 1/2\gamma$
which can take any value in $]0,1[$ even though the spectrum is always
absolutely continuous.\\

\noindent Another interesting case is $t_k =(q-1)/q^{(k+1)}$ (geometrical
scaling)
with $q>1$, for which $||H|| =1$.

\noindent (i) For $1<q\leq2$, the spectrum is nothing but the $q-$adic
decomposition of
real numbers in the interval $[-1,+1]$. It is therefore gapless and absolutely
continuous.

\noindent (ii) For $q>2$, the image of $E$ is a Cantor set of zero Lebesgue
measure,
constructed by removing the central interval of width $2(1-2/q)$ in the
interval
$[-1,+1]$ and repeating the operation on each of the intervals left. The
spectrum is a monofractal set with a Hausdorf dimension $D_H=\ln 2/\ln q$
\cite{Kahane}. The spectral measure is the Cantor one and gives the same weight
to each subinterval.  Note that the classical tryadic Cantor set is
obtained for
$q=3$.

\noindent For such Cantor spectra, it is shown in ref.\cite{KPG} that the
temporal correlation function decays as $C(t) \sim t^{-D_2}$, where $D_2$
is the
correlation dimension of the spectral measure ({\it i. e.} of the local density
of states). In this example, one has: $D_2=1$ for $1<q\leq2$, since the
spectrum
is absolutely continuous, and  $D_2=D_H=\log 2/\log q$ for $q>2$  since the
spectrum is then a monofractal set. It is important to consider $C$ because
the behaviour of $P$ is much
more complex. In particular, $P$ is sensitive to the nature of $q$.
Indeed, it is shown in \cite{Salem} that:

%%%%%%%%%%%%%%%
\begin{equation}
\lim_{s\rightarrow \infty} P(s)=0\\
\Leftrightarrow q \notin S\setminus \{2\}
\mbox{ , }
\end{equation}
%%%%%%%%%%%%%%%

\noindent where $S$ denotes the set of algebraic integer numbers defined by
Pisot and Vijayaraghavan
\cite{Pisot,Pisot2}.\\
Note that the functional relation:

%%%%%%%%%%%%%%%
\begin{equation}
\label{functional}
\forall q, \hspace{1.ex} P(qs)=\cos^2(s(q-1)) P(s)
\mbox{ , }
\end{equation}
%%%%%%%%%%%%%%%

\noindent allows us to exactly determine $P$ for $q=2$ since:

%%%%%%%%%%%%%%%
\begin{equation}
P(2s)=\cos^2(s) P(s)\Leftrightarrow P(s)=\sin^2(s)/s^2
\mbox{ . }
\end{equation}
%%%%%%%%%%%%%%%

\noindent According to the identity (\ref{beta}), one obtains a diffusion
exponent $\beta =0$
for any $q>1$ whereas the spectrum can be either absolutely continous or
singular continuous. In addition, one can show  that $L^2(s) \sim \ln{(s)}$,
with a criterion similar to the one given in eq. (\ref{definition}).

\vspace{.5cm}

To conclude, these toy models defined on infinite dimensional hypercubes allows
us to carefully analyze the possible relationships between the spectral measure
and the diffusion exponents. The first class of hamiltonians (algebraic scaling
of the hopping terms), shows that it is possible to face an absolutely
continuous spectrum and an anomalous diffusion with a $\beta$ exponent that can
take any value between $0$ and $1$. On the other hand, the second case
(geometrical scaling of the hopping parameters), displays a zero $\beta$
exponent whereas the spectrum can be either absolutely or singular
continuous. Finally, we emphasize upon the importance of these exponents
in transport properties, especially in quasicrystals, where they should be
responsible for the anomalous behaviour of their conductivity.

\vspace{1.5cm}

{\large \bf Appendix:  proof of eq.~(\ref{Pest})}

\vspace{.2cm}

Let us consider $K>0$ large enough so that $t/2k^{\gamma}\leq t_k \leq
2t/k^{\gamma}$ for
$k\geq K$. Then choose $s_0 >0$ large enough so that $s_0 t/K^{\gamma} \leq
\pi/2 \leq s_0 t/(K-1)^{\gamma}$. For $s \geq s_0$ let $K_1 \geq K$ be such
that
$st/2K_1^{\gamma} <\pi/2\leq st/2(K_1-1)^{\gamma}$. Then

$$
\ln{P(s)} \leq
 \sum_{k=K_1}^{\infty} \ln{\cos^2{st/2k^{\gamma}}}
\mbox{ , }
$$

\noindent If one sets $x_k = k\left( 2/st \right)^{1/\gamma}$, the
right hand side is dominated by an integral of the form

$$
\ln{P(s)} \leq
 (ts/2)^{1/\gamma}
  \int_{(2/\pi)^{1/\gamma}+ O(s^{1/\gamma})}^{\infty}
   dx \ln{\cos^2(1/x^{\gamma})}
\mbox{ , }
$$

\noindent for $s>s_0$, leading to eq.~(\ref{Pest}).


\begin{references}{}

\bibitem{KPG} R. Ketzmerick, G. Petschel and T. Geisel, {\em Phys. Rev.
Lett.}, {\bf 69}, 695, (1992).

\bibitem{KKKG} R. Ketzmerick, {\it et al.}, {\em Phys. Rev. Lett.},
{\bf 79}, 1959, (1997).

\bibitem{Guarneri} I. Guarneri, {\em Europhys. Lett.}, {\bf 21}, 729, (1993).

\bibitem{Zhong} J. X. Zhong and R. Mosseri, {\em J. Phys.: Condens. Matter},
{\bf 7}, 8383, (1995).

\bibitem{Suto} A. S\"ut\"o, {\em J. Stat. Phys.}, {\bf 56}, 525, (1989).

\bibitem{BTS} J. Bellissard, B. Iochum, E. Scoppola and D. Testard, {\em Comm.
Math. Phys.}, {\bf 125}, 527-543, (1989).

\bibitem{Be88} J. Bellissard, in {\em Number Theory and Physics} pp. 140-150,
Les Houches Mars 89, Springer Proc. in Physics, vol.47, J.M. Luck, P. Moussa
\& M. Waldschmidt Eds., (1989).

\bibitem{BBG} J. Bellissard, A. Bovier and J.M. Ghez, {\em Comm. Math. Phys.},
{\bf 135}, 379-399, (1991).

\bibitem{BovGhe} A. Bovier and J.M. Ghez, {\em J. Phys.}, {\bf A28}, 2213-2224,
(1995).

\bibitem{SimKnill} A. Hof, O. Knill and B. Simon, {\em Comm. Math. Phys.}, {\bf
174}, 149-159, (1995).

\bibitem{Hofstadter}  D. R. Hofstadter, {\em Phys. Rev.}, {\bf B 14}, 2239
(1976).

\bibitem{XX} Y. Last, {\em Comm. Math. Phys.}, {\bf 164}, 421-432, (1994).

\bibitem{XX2} H. Hiramoto and M. Kohmoto, {\em Phys. Rev.}, {\bf B40},
8225-8234, (1989).

\bibitem{Sire2} C. Sire, {\em Europhys. Lett.}, {\bf 10}, 483-488, (1989).

\bibitem{Benza} V. G. Benza and C. Sire {\em Phys. Rev.},
{\bf B44}, 10343-10345, (1991).

\bibitem{Sire3} B. Passaro, C. Sire, and V. G. Benza {\em Phys. Rev.},
{\bf B46}, 13751-13755, (1992).

\bibitem{GuaSchuba} I. Guarneri, H. Schulz-Baldes, mp\_arc 98-382, mp\_arc
98-457.

\bibitem{SchuBar} H. Schulz-Baldes and J. M. Barbaroux, {\em Anomalous quantum
transport in presence of self-similar spectra}, submitted to Ann. IHP, (1998).

\bibitem{BBM} J. Bellissard, D. Bessis and P. Moussa, {\em Phys. Rev. Lett.},
{\bf 49}, 701-704, (1982).

\bibitem{Barnsley} M. Barnsley, J. Geronimo and Harrington {\em Comm. Math.
Phys.}, {\bf 99}, 303-317, (1985).

\bibitem{BeSchuba} H. Schulz-Baldes and J. Bellissard, {\em Rev. Math. Phys.},
{\bf 10}, 1-46, (1998).

%\bibitem{Geisel} T. Geisel, R. Ketzmeric, and G. Petschel, Phys. Rev.
%Lett., {\bf 66}, 1651-1654, (1991).

\bibitem{FujiRo} T. Fujiwara and S. Roche, {\em Phys. Rev.}, {\bf B58},
11338-11396, (1998).

\bibitem{remark} $\beta(E)=0$ does not necessarily implies that $L_E(t)/t$
converges to a positive constant as $t\rightarrow \infty$. Because
$\beta (E)$ is the infimum of the $\gamma$'s such that
$\int_1^{\infty} dt L(t)/t^{1+\gamma} < \infty$ (see eq.~\ref{definition}). A
counter example can be found in \cite{Simon2}.

\bibitem{Simon2} B. Simon, {\em Comm. Math. Phys.}, {\bf 134}, 209-213, (1990).

\bibitem{Pastur} A. Khorunzhy and L. Pastur, {\em Comm. Math. Phys.},
{\bf 153}, 605-646, (1993); F. Wegner, {\em Phys. Rev.}, {\bf B19}, 783-792,
(1979).

\bibitem{BZM} J.X. Zhong, J. Bellissard and R. Mosseri, {\em J. Phys: Condens.
Matter}, {\bf 7}, (1995), 3507-3514.

\bibitem{Kahane} J. P. Kahane and R. Salem, {\em Ensembles parfaits et s\'eries
trigonom\'etriques} p.72, Hermann, Paris, (1963).

\bibitem{Salem} R. Salem, {\em Set of uniqueness and sets of multiplicity},
Trans. Amer. Math. Soc.,{ \bf 54}, 1943; Trans. Amer. Math. Soc., {\bf 56},
1944; Trans. Amer. Math. Soc., {\bf 63}, (1948).

\bibitem{Pisot} C. Pisot, {\em La r\'epartition modulo $1$ et les nombres
alg\'ebriques}, Annali di Pisa, {\bf 7}, (1938).

\bibitem{Pisot2}  C. Pisot, {\em Sur une famille remarquable d'entiers
al\-g\'e\-bri\-ques formant un ensemble ferm\'e}, Colloque sur la th\'eorie des
nombres, Bruxelles, (1955).

\end{references}
\end{document}